\documentclass[12pt]{iopart}

\usepackage{times}
\usepackage{epsfig}
\usepackage{graphics}
\usepackage{cite}
\usepackage{fleqn}
\usepackage{amssymb}
\usepackage{rotating}

\newcommand{\re}{\mathop{\mathrm{Re}}}

\newcommand{\D}{\mathop{\mathrm{d}}}
\newcommand{\I}{\mathop{\mathrm{i}}}
\newcommand{\E}{\mathop{\mathrm{e}}}

\begin{document}

\title[Statistical and coherence properties of radiation from X-ray free
electron lasers]{Statistical and coherence properties of radiation from
X-ray free electron lasers\footnote{\normalsize \sl DESY Print DESY 09-224,
Submitted to New Journal of Physics.}}

\author{E.L. Saldin, E.A. Schneidmiller, M.V. Yurkov}

\address{Deutsches Elektronen-Synchrotron DESY, 22607 Hamburg, Germany}
\ead{mikhail.yurkov@desy.de}

\begin{abstract}

We describe statistical and coherence properties of the radiation from x-ray
free electron lasers (XFEL). It is shown that the X-ray FEL radiation
before saturation is described with gaussian statistics. Particularly important
is the case of the optimized X-ray FEL, studied in detail. Applying similarity
techniques to the results of numerical simulations allowed us to find universal
scaling relations for the main characteristics of an X-ray FEL operating in the
saturation regime: efficiency, coherence time and degree of transverse
coherence. We find that with an appropriate normalization of these quantities,
they are functions of only the ratio of the geometrical emittance of the
electron beam to the radiation wavelength. Statistical and coherence properties
of the higher harmonics of the radiation are highlighted as well.

\end{abstract}

\pacs{41.60.Cr, 52.59.Rz, 42.25.Kb, 42.55.Vc}

\maketitle

\section{Introduction}

Single pass free electron laser (FEL) amplifiers starting from shot noise in
the electron beam have been intensively developed during the last
decades\footnote{Following the terminology of quantum lasers (amplified
spontaneous emission, ASE), the term "self amplified
spontaneous emission (SASE)" in connection with an FEL amplifier, starting from
shot noise, started to be used in \cite{boni-sase}. Note that
this essentially quantum terminology does not reflect physical properties of
the device. In fact, free electron laser belongs to a separate class of vacuum
tube devices, and its operation is completely described in terms of classical
physics (see \cite{book} for more detail).}.
An origin for this development was an idea born in the early eighties to extend
the operating wavelength range of FELs to the vacuum ultraviolet (VUV) and x-ray
bands \cite{ks-sase,dks-sase,pellegrini-murphy-sase}. Significant efforts have
been invested into the development of high brightness injectors, beam formation
systems, linear accelerators, and undulators. The result was rapid extension
of the wavelength range from infrared to hard
x-rays
\cite{pellegrini-sase,leutl-sat,visa,ttf-sat-prl,ttf-sat-epj,vuvfel-exp,flash-nat-phot,scss-nat-phot,lcls-epac,lcls-fel09}.
The first dedicated user facility FLASH at
DESY in Hamburg is in operation since 2005 and provides wavelength range from
6.5~nm to 50 nm \cite{flash-njp}. LCLS at Stanford has been recently
commissioned and delivers radiation in the 0.15 nm - 1.5 nm wavelength range
\cite{lcls-fel09}. The two other dedicated facilities that are under
construction at the moment, the European XFEL, and SCSS at Spring-8
\cite{scss-cdr,euro-xfel-tdr}.

The high gain FEL amplifier starting from the shot noise in the electron beam
is a very simple device. It is a system consisting of a relativistic electron
beam and an undulator. The FEL collective instability in the electron beam
produces an exponential growth (along the undulator) of the modulation of the
electron density on the scale of undulator radiation wavelength. The initial
seed for the amplification process are fluctuations of the electron beam
current. Since shot noise in the electron beam is a stochastic process, the
radiation  produced by a SASE FEL possesses stochastic features as well. Its
properties are naturally described in terms of statistical optics using notions
of probability density distribution functions of the fields and intensities,
correlation functions, notions of coherence time, degree of coherence, etc.

Development of the theoretical description of the coherence properties of the
radiation from SASE FEL has spanned more than twenty years
(see \cite{1d1,1d2,1d3,1d4,1d5,1d6,1d7,ssy-sb,ssy-flash-femto,sam-kr-1,sam-kr-2,
sam-kr-3,sam-kr-4,
trcoh-oc,trcoh-nima,coherence-oc,coherence-anal-oc,kjkim-review}.
This subject is rather complicated, and it is worth mentioning that theoretical
predictions agree well with recent experimental results
\cite{ttf-sat-prl,ttf-sat-epj,vuvfel-exp,flash-nat-phot,flash-statistics,flash-coherence,leutl-frog1,leutl-frog2}.
Some averaged output characteristics of SASE FEL in framework of
the one-dimensional model have been obtained in \cite{1d1,1d2}. An approach for
time-dependent numerical simulations of SASE FEL has been developed in
\cite{1d3,1d4}. Realization of this approach allowed one to obtain some
statistical properties of the radiation from a SASE FEL operating in linear and
nonlinear regime \cite{1d5,1d6}. A comprehensive study of the statistical
properties of the radiation from the SASE FEL in the framework of the same model
is presented in \cite{1d7}. It has been shown that a SASE FEL operating in
the linear regime is a completely chaotic polarized radiation source described
with gaussian statistics. Short-pulse effects (for pulse durations comparable
with the coherence time) have been studied in
\cite{1d4,ssy-sb,ssy-flash-femto}. An important practical result was prediction
of the significant suppression of the fluctuations of the radiation intensity
after a narrow-band monochromator for the case of SASE FEL operation in the
saturation regime \cite{ssy-sb}. Statistical description of the chaotic
evolution of the radiation from SASE FEL has been presented in
\cite{sam-kr-1,sam-kr-2}.

The first analytical studies of the problem of transverse coherence relate to
the late eighties \cite{sam-kr-3,sam-kr-4}. Later on more detailed studies have
been performed \cite{trcoh-oc}. The problem of start-up from the shot noise has
been studied analytically and numerically for the linear stage of amplification
using an approach developed in \cite{sam-kr-4}. It has been found that the
process of formation of transverse coherence is more complicated than that
given by a naive physical picture of transverse mode selection. Namely, in the
case of perfect mode selection the degree of transverse coherence is defined by
the interdependence of the longitudinal and transverse coherence. Comprehensive
studies of the evolution of transverse coherence in the linear and nonlinear
regime of SASE FEL operation have been performed in
\cite{trcoh-nima,coherence-oc,coherence-anal-oc}. It has been found that the
coherence time and the degree of transverse coherence reach maximum values in
the end of the linear regime. Maximum brilliance of the radiation is achieved
in the very beginning of the nonlinear regime which is also referred as a
saturation point \cite{coherence-oc}. Output power of the SASE FEL grows
continuously in the nonlinear regime, while the brilliance drops down after
passing saturation point.

\section{Operation of an FEL amplifier}

A single-pass FEL amplifier starting from the shot noise in the
electron beam seems similar to the well known undulator insertion device:
in both cases radiation is produced during single pass of the electron beam
through the undulator. To reveal principal differences, we first recall the
properties of incoherent radiation. Radiation within the cone of half angle
$\theta_{\mathrm{con}} = \sqrt{1 + K^{2}}/(\gamma\sqrt{N_{\mathrm{w}}})$ has
relative spectral bandwidth $\Delta\lambda /\lambda \simeq 1/N_{\mathrm{w}}$
near the resonance wavelength $\lambda =
(\lambda_{\mathrm{w}}/2\gamma^{2})(1+K^{2})$. Here $\lambda_{\mathrm{w}}$ is
the undulator period, $N_{\mathrm{w}}$ is the number of undulator periods,
$\gamma $ is the relativistic factor, $K = e \lambda_{\mathrm{w}} H_{\mathrm{w}}
/ 2 \pi m c^2$ is the undulator parameter, $H_{\mathrm{w}}$ is the rms undulator
field, and $m$ and $e$ are the electron mass and charge, respectively. Radiation
energy emitted by a single electron in the central cone is $E_\mathrm{e}
\simeq 4\pi ^2 e^2 K^2 A_{\mathrm{JJ}}^{2} /[\lambda(1+K^{2})]$. Here
$A_{\mathrm{JJ}} = 1$ and $A_{\mathrm{JJ}} = [J_{0}(Q) - J_{1}(Q)]$ for a
helical and a planar undulator, respectively, $J_{n}(Q)$ is a Bessel function
of $n$th order, and $Q = K^{2}/2/(1 + K^{2})$. Wavepackets emitted by different
electrons are not correlated, thus radiated power of the electron bunch with
current $I$ is just the radiation energy from a single electron multiplied by
the electron flux $I/e$:

\begin{equation}
W_\mathrm{incoh} \simeq
\left[\frac{4\pi ^2 eI}{\lambda }\right]
\left[ \frac{K^{2}}{1+K^{2}}\right]A_{\mathrm{JJ}}^{2} \ .
\label{eq:power-incoherent}
\end{equation}

\noindent In the free electron laser, electron beam density is modulated by
the period of resonance wavelength $\lambda $. Let us consider a model case of an
electron beam with the gaussian distribution of the current density with rms
width $\sigma $, and an axial modulation $I(z) = I [1+a_{\mathrm{in}}\cos(2\pi
z/\lambda )]$. Total power radiated by a modulated electron beam has been
derived in \cite{frog-diag-2005}:

\begin{equation}
W_\mathrm{coh} =
\left[\frac{\pi^{2}a^{2}_{\mathrm{in}} I^2 }{2c}\right]
\left[\frac{K^{2}}{1 + K^{2}}\right]A^{2}_{\mathrm{JJ}}N_{\mathrm{w}}
F(N) \ ,
\label{eq:power-modulated}
\end{equation}

\noindent where

\begin{equation}
F(N) = \frac{2}{\pi}\left[\arctan\left(\frac{1}{2N}\right)
+ N\ln\left(\frac{4N^{2}}{4N^{2}+1}\right)\right] \ ,
\label{eq:f}
\end{equation}

\begin{figure}[tb]
\begin{center}
\epsfig{file=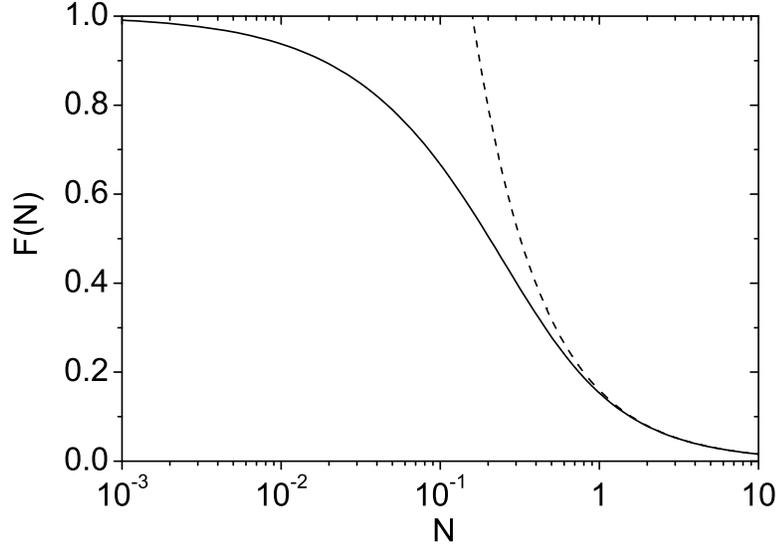,width=0.75\textwidth}
\end{center}
\caption{
Plot of the universal function F (N) given by eq. (\ref{eq:f}).
Dashed curve shows an asymptote for a wide
electron beam $F = 1/(2\pi N)$.
}
\label{fig:pcohu}
\end{figure}

\noindent $N = k\sigma^{2}/L_{\mathrm{w}}$  is the Fresnel number, $k = 2\pi
/\lambda $, and $L_{\mathrm{w}} = N_{\mathrm{w}}\lambda_{\mathrm{w}}$ is the
undulator length. In Fig.~\ref{fig:pcohu} we present the plot of the universal
function $F(N)$. It exhibits a simple behavior in the limits of large and
small values of Fresnel number: $F(N) \to 1/(2\pi N)$ for $N \to \infty$, and
$F(N) \to 1$ for $N \to 0$.

Analysis of expressions (\ref{eq:power-incoherent}) and
(\ref{eq:power-modulated}) tells us that incoherent radiation power corresponds
to the radiation power of the modulated electron beam with effective modulation
amplitude of $a_{\mathrm{in}} \sim 1/ \sqrt{N_{\mathrm{w}}I\lambda /(ec)}$.
Note that $N_{\mathrm{w}}I\lambda /(ec)$ is the number of electrons on the
slippage length $N_{\mathrm{w}}\lambda $. Now we have quantitative answer to
the question: how much the FEL radiation power exceeds the power of incoherent
undulator radiation? In the free electron laser an amplitude of the electron
beam density modulation reaches values of about unity, and the ratio of the
 radiation powers (coherent to incoherent) is a factor of about the
number of electrons per slippage length.

Enhancement of the beam modulation in the free electron laser occurs due to
the radiation-induced collective instability. When an electron beam traverses
an undulator, it emits radiation at the resonance wavelength. The
electromagnetic wave is always faster than the electrons, and a resonant
condition occurs such that the radiation slips a distance $\lambda$ relative to
the electrons after one undulator period. The fields produced by the moving
charges in one part of the electron bunch react to moving charges in another
part of the bunch leading to a growing concentration of particles wherever a
small perturbation starts to occur. Electron bunches with very small
transverse emittance (of about radiation wavelength) and high peak current (of
about a few kA) are needed for the operation of short wavelength FELs.

The description of high gain FEL amplification refers to the class of
self-consistent problems for which the field equations and equations of motion
must be solved simultaneously. Characteristics of the amplification process
can be obtained with a combination of analytical techniques and simulations
with time dependent FEL simulation codes (see \cite{book,kjkim-review} and
references therein). The amplification process in the SASE FEL is triggered by
the shot noise in the electron beam, then it passes the stage of exponential
growth (also called the high gain linear regime), and finally enters saturation
stage when the beam density modulation approaches unity. In the linear high-gain
limit the radiation emitted by the electron beam in the undulator can be
represented as a set of self-reproduced beam radiation modes
\cite{moore-modes}:

\begin{equation}
\tilde{E} = \int \mathrm{d}\omega
\exp [i\omega (z/c-t)]
 \times \sum \limits _{n,k}
A_{nk}(\omega , z) \Phi_{nk}(r,\omega )
\exp [ \Lambda _{nk}(\omega )z + in\phi ]
\label{eq:modes}
\end{equation}

\noindent described by the eigenvalue $\Lambda _{nk}(\omega )$ and the field
distribution eigenfunction $\Phi _{nk}(r,\omega )$. Here $\omega = 2\pi
c/\lambda $ is the frequency of the electromagnetic wave. At a sufficient
undulator length the fundamental mode (having maximum real part of the
eigenvalue) begins to be the main contribution to the total radiation power.
From a practical point of view, it is important to find an absolute minimum of
the gain length $L_{\mathrm{g}} = 1/\re(\Lambda _{00})$ corresponding to the
optimum focusing beta function. In the case of negligible space charge and
energy spread effects (which is true for XFELs) the solution of the
eigenvalue equation for the field gain length of the fundamental mode
$L_{\mathrm{g}}$ and optimum beta function $\beta_{\mathrm{opt}}$ are
well approximated by\footnote{General fitting expression including
energy spread effects can be found in \cite{xfel-fit}.}:

\begin{eqnarray}
L_{\mathrm{g}} & \simeq &
1.67 \left(\frac{I_A}{I} \right)^{1/2}
\frac{(\epsilon_n \lambda_{\mathrm{w}})^{5/6}}
{\lambda ^{2/3}} \ \frac{(1+K^2)^{1/3}}{K A_{JJ}}  \ ,
\nonumber \\
\beta_{\mathrm{opt}} & \simeq & 11.2 \left(\frac{I_A}{I} \right)^{1/2}
\frac{\epsilon_n^{3/2}
\lambda_{\mathrm{w}}^{1/2}}
{\lambda K A_{JJ}}  \ .
\label{eq:lg}
\end{eqnarray}

\noindent Here $\epsilon _{\mathrm{n}}$ is normalized emittance, and $I_{A} =
mc^3/e \simeq 17$~kA is Alfven's current.
Dimensionless FEL equations are normalized using the gain parameter $\Gamma $
and the efficiency parameter $\bar{\rho }$ \cite{book}:

\begin{eqnarray}
\Gamma & = & \left[ \frac{I}{I_{\mathrm{A}}} \frac{8 \pi ^2 K^2
A_{\mathrm{JJ}}^2}{\lambda \lambda _\mathrm{w} \gamma ^3}
\right]^{1/2} \ ,
\nonumber \\
\bar{\rho} & = & \frac{\lambda _\mathrm{w}\Gamma }{4 \pi } \ .
\label{eq:rho-3d}
\end{eqnarray}

\noindent Analysis of the dimensionless FEL equations tells us that the
physical parameters describing operation of the optimized FEL (\ref{eq:lg}),
the diffraction parameter $B$ and the parameter of betatron oscillations
$\hat{k}_{\beta }$, are only functions of the parameter $\hat{\epsilon } = 2\pi
\epsilon/\lambda $\cite{book,coherence-oc,eigen-general}:

\begin{eqnarray}
B & = & 2 \Gamma \sigma^2 \omega/c \simeq 12.5 \times \hat{\epsilon }^{5/2} \ ,
\nonumber \\
\hat{k}_{\beta} & = & 1/(\beta \Gamma )
\simeq 0.158 \times \hat{\epsilon }^{-3/2} \ .
\label{eq:reduced-parameters}
\end{eqnarray}

\noindent Note that Eqs.~(\ref{eq:lg}) and (\ref{eq:reduced-parameters}) are
accurate in the range $1 < \hat{\epsilon } < 5$.

The diffraction parameter $B$ directly relates to diffraction effects
and the formation of transverse coherence. If diffraction expansion of the
radiation on a scale of the field gain length is comparable with the transverse
size of the electron beam, we can expect a high degree of transverse coherence.
For this range of parameters the value of the diffraction parameter is small.
If diffraction expansion of the radiation is small (which happens at large
values of the diffraction parameter) then we can expect significant degradation
in the degree of transverse coherence. This effect occurs simply because
different parts of the beam produce radiation nearly independently. In terms of
the radiation expansion in the eigenmodes (\ref{eq:modes}) this range of
parameters corresponds to the degeneration of modes \cite{book}. Diffraction
parameter for an optimized XFEL exhibits strong dependence on the parameter
$\hat{\epsilon }$ (see eq.~(\ref{eq:reduced-parameters})), and we can expect
that the degree of transverse coherence should drop rapidly with the increase
of the parameter $\hat{\epsilon }$.

\section{Definitions of the statistical properties of radiation}

We describe radiation fields generated by a SASE FEL in terms of statistical
optics \cite{goodman}. Longitudinal and transverse coherence are described in
terms of correlation functions. The first order time correlation function,
$g_1(t,t')$, is defined as:

\begin{equation}
g_1(\vec{r},t-t')  =
\frac{\langle \tilde{E}(\vec{r},t)\tilde{E}^*(\vec{r},t')\rangle }
{\left[\langle \mid\tilde{E}(\vec{r},t)\mid^2\rangle
\langle \mid\tilde{E}(\vec{r},t')\mid^2\rangle \right]^{1/2}} \ .
\label{def-corfunction}
\end{equation}

\noindent For a stationary random process the time correlation functions are
dependent on only one variable, $\tau = t - t'$. The coherence time is
defined as \cite{mandel,book}:

\begin{equation}
\tau_{\mathrm{c}} = \int \limits^{\infty}_{-\infty}
| g_1(\tau) |^2 \D\tau \ .
\label{coherence-time-def}
\end{equation}

    The transverse coherence properties of the radiation are described in terms
of the transverse correlation functions.  The first-order transverse
correlation function is defined as

\begin{displaymath}
\gamma_1 (\vec{r}_{\perp}, \vec{r}\prime _{\perp}, z, t) = \frac{
\langle \tilde{E} (\vec{r}_{\perp}, z, t)
\tilde{E}^{*} (\vec{r}\prime _{\perp}, z, t) \rangle }
{ \left[ \langle |\tilde{E} (\vec{r}_{\perp}, z, t) |^2 \rangle
\langle |\tilde{E} (\vec{r}\prime _{\perp}, z, t) |^2 \rangle \right]^{1/2}}
\ ,
\end{displaymath}

\noindent where $\tilde{E}$ is the slowly varying amplitude of the
amplified wave,
$E =
\tilde{E} (\vec{r}_{\perp}, z, t) \E^{\I \omega_0 (z/c - t)}
+ {\mathrm{C.C.}}$
We consider the model of a stationary random
process, meaning that $\gamma_1$ does not depend on time. Following
ref.~\cite{coherence-oc}, we define the degree of transverse coherence as:

\begin{equation}
\zeta =
\frac
{
\int |\gamma _1 (\vec{r}_{\perp},\vec{r}\prime _{\perp})|^{2}
I(\vec{r}_{\perp})
I(\vec{r}\prime _{\perp})
\D\vec{r}_{\perp}
\D\vec{r}\prime _{\perp}
}
{
[\int I(\vec{r}_{\perp}) \D\vec{r}_{\perp}]^{2}
} \ ,
\label{eq:def-degcoh}
\end{equation}

\noindent where $I \propto |\tilde{E}|^2$ is the radiation intensity.

An important figure of merit of the radiation source is the degeneracy
parameter $\delta $, the number of photons per mode (coherent state). Note
that when $\delta \gg 1$, classical statistics are applicable, while a quantum
description of the field is necessary as soon as $\delta$ is comparable to (or
less than) one. Using the definitions of the coherence time
(\ref{coherence-time-def}) and of the degree of transverse coherence
(\ref{eq:def-degcoh}) we define  the degeneracy parameter as

\begin{equation}
\delta = \dot{N}_{ph} \tau_{\mathrm{c}} \zeta \ ,
\label{degeneracy-def}
\end{equation}

\noindent where $\dot{N}_{ph}$ is the photon flux. Peak brilliance of the
radiation from an undulator is defined as a transversely coherent spectral flux:

\begin{equation}
B_r =
\frac{\omega \D \dot{N}_{ph}}{\D \omega} \
\frac{\zeta}{\left(\lambda/2\right)^2} =
\frac{4\sqrt{2} c\delta }{\lambda^3}  \ .
\label{eq:bril-1}
\end{equation}

\noindent When deriving right-hand term of the equation we used the fact
that the spectrum shape of SASE FEL radiation in a high-gain linear regime and
near saturation is close to Gaussian \cite{book}. In this case
the rms spectrum bandwidth $\sigma_{\omega}$ and coherence time obey the
equation $\tau_{\mathrm{c}} = \sqrt{\pi}/\sigma_{\omega}$.

\section{Probability distributions of the radiation fields and intensities}

\begin{figure}[tb]
\begin{center}
\epsfig{file=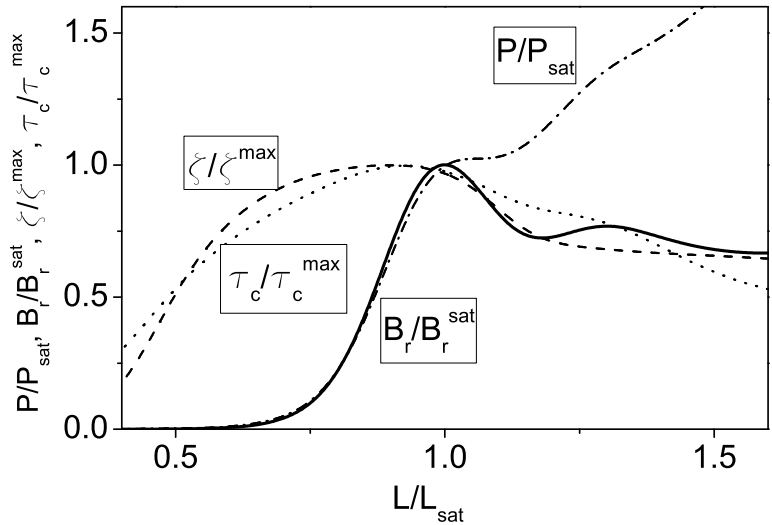,width=0.75\textwidth}
\end{center}
\caption{
Evolution of main characteristics of SASE FEL along the undulator: brilliance
(solid line), radiation power (dash-dotted line), degree of transverse coherence
(dashed line), and coherence time (dotted line). Brilliance and radiation power
are normalized to saturation values. Coherence time and degree of transverse
coherence are normalized to the maximum values. Undulator length is normalized
to saturation length. The plot has been derived from the parameter set
corresponding to $\hat{\epsilon } = 1$. Calculations have been performed with
the simulation code FAST \cite{fast}.}

\label{fig:em1-delta}
\end{figure}

\begin{figure}[tb]

\includegraphics[width=0.5\textwidth]{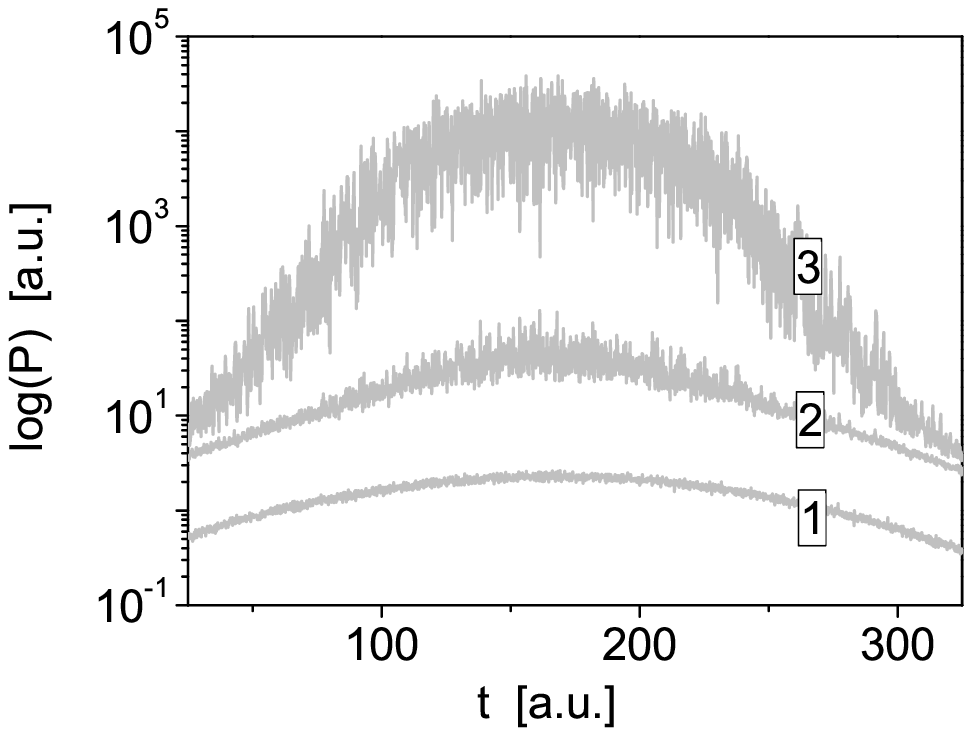}
\includegraphics[width=0.5\textwidth]{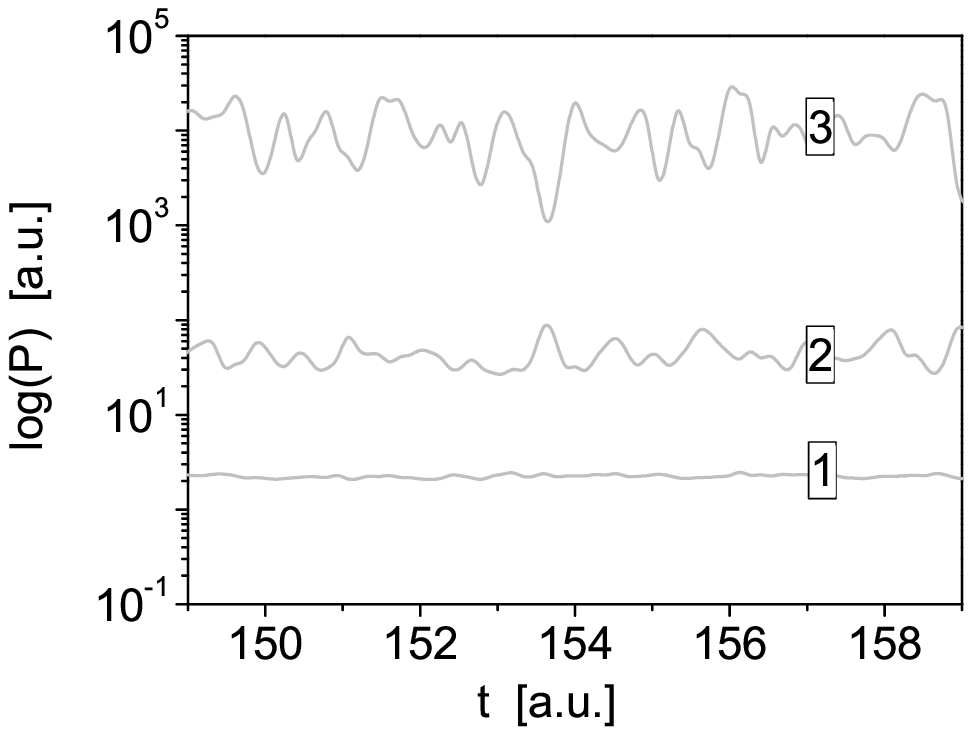}

\caption{Temporal structure of the radiation
pulse at different undulator lengths. Indexes 1, 2, and 3 correspond to the
undulator length of $0.5L_\mathrm{g}$, $5L_\mathrm{g}$, and $10L_\mathrm{g}$,
respectively. The plot in the right column represents the enlarged fraction of
the plot in the left column. Calculations have been performed with
the simulation code FAST \cite{fast}. }

\label{fig:temporal-sase1}
\end{figure}

\begin{figure}[tb]

\includegraphics[width=0.5\textwidth]{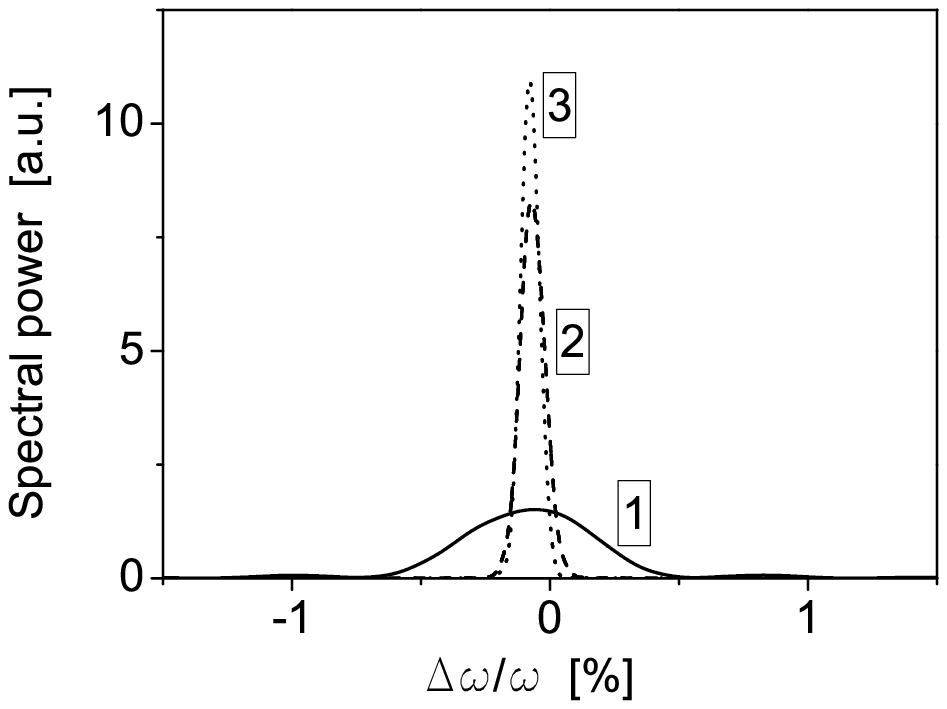}
\includegraphics[width=0.5\textwidth]{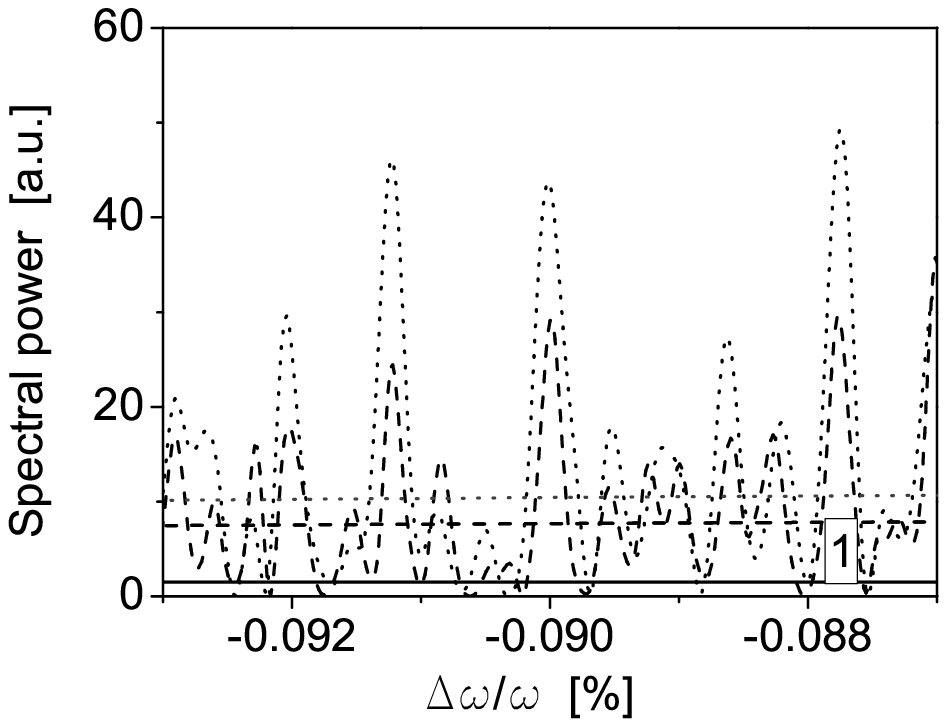}

\caption{Spectral structure of the radiation
pulse at different undulator length. Solid, dashed, and dotted line correspond
to the undulator length of $0.5L_\mathrm{g}$, $5L_\mathrm{g}$, and
$10L_\mathrm{g}$, respectively. The plot on the left-hand side shows only
the envelope of the radiation spectrum. The plot on the right-hand side
represents an enlarged fraction of the radiation spectrum. Calculations have
been performed with the simulation code FAST \cite{fast}. }

\label{fig:spectral-sase1}
\end{figure}

\begin{figure}[tb]

\includegraphics[width=0.33\textwidth]{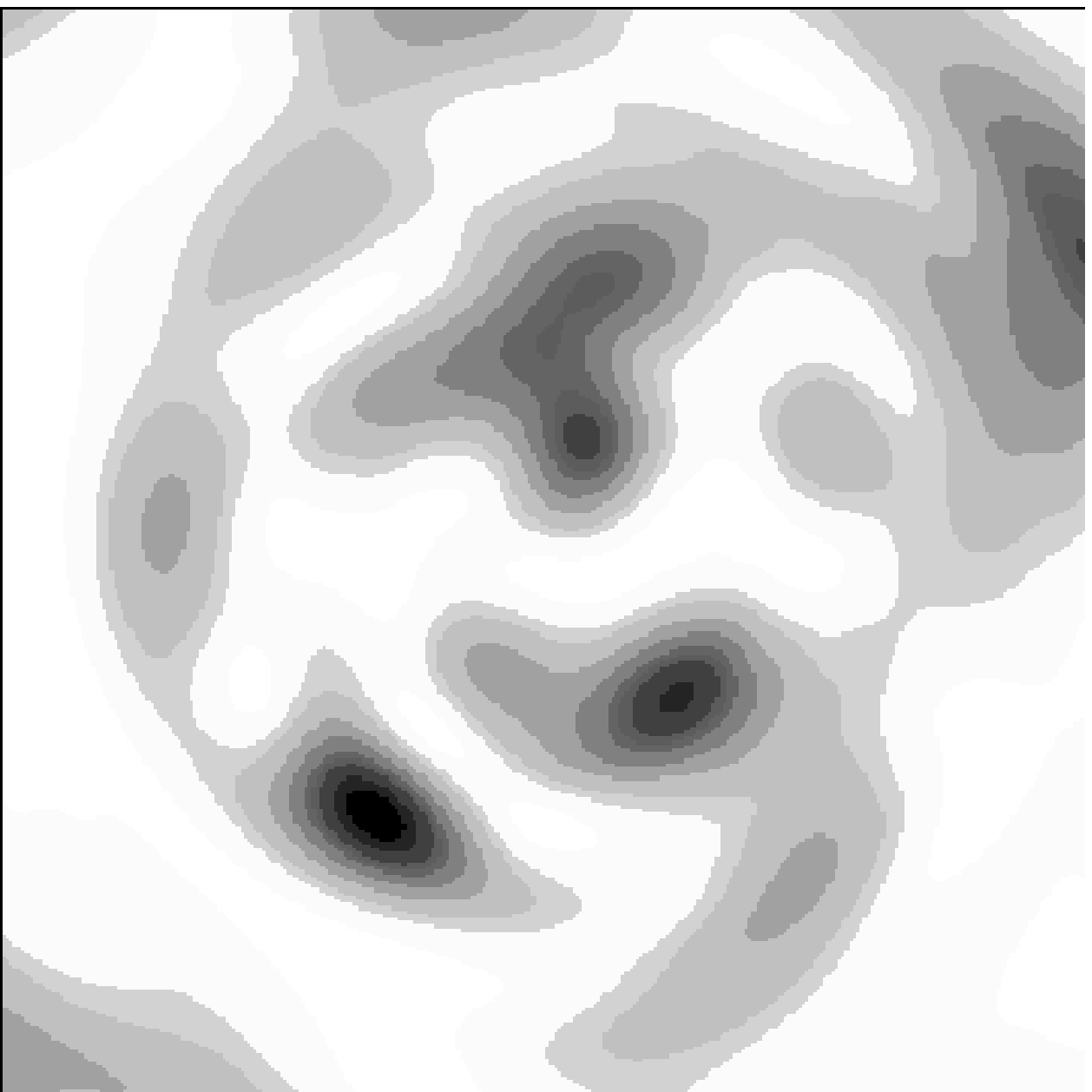}
\includegraphics[width=0.33\textwidth]{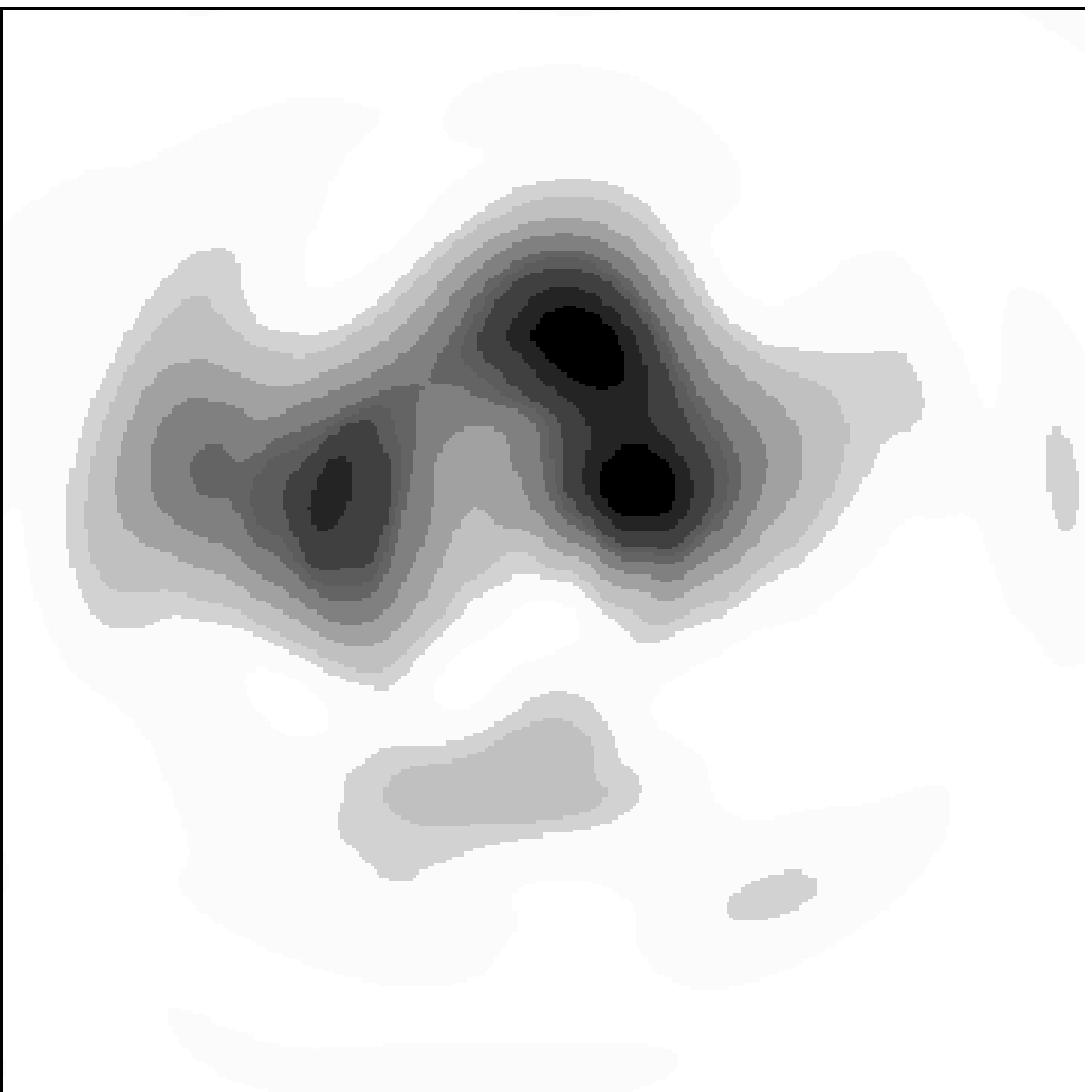}
\includegraphics[width=0.33\textwidth]{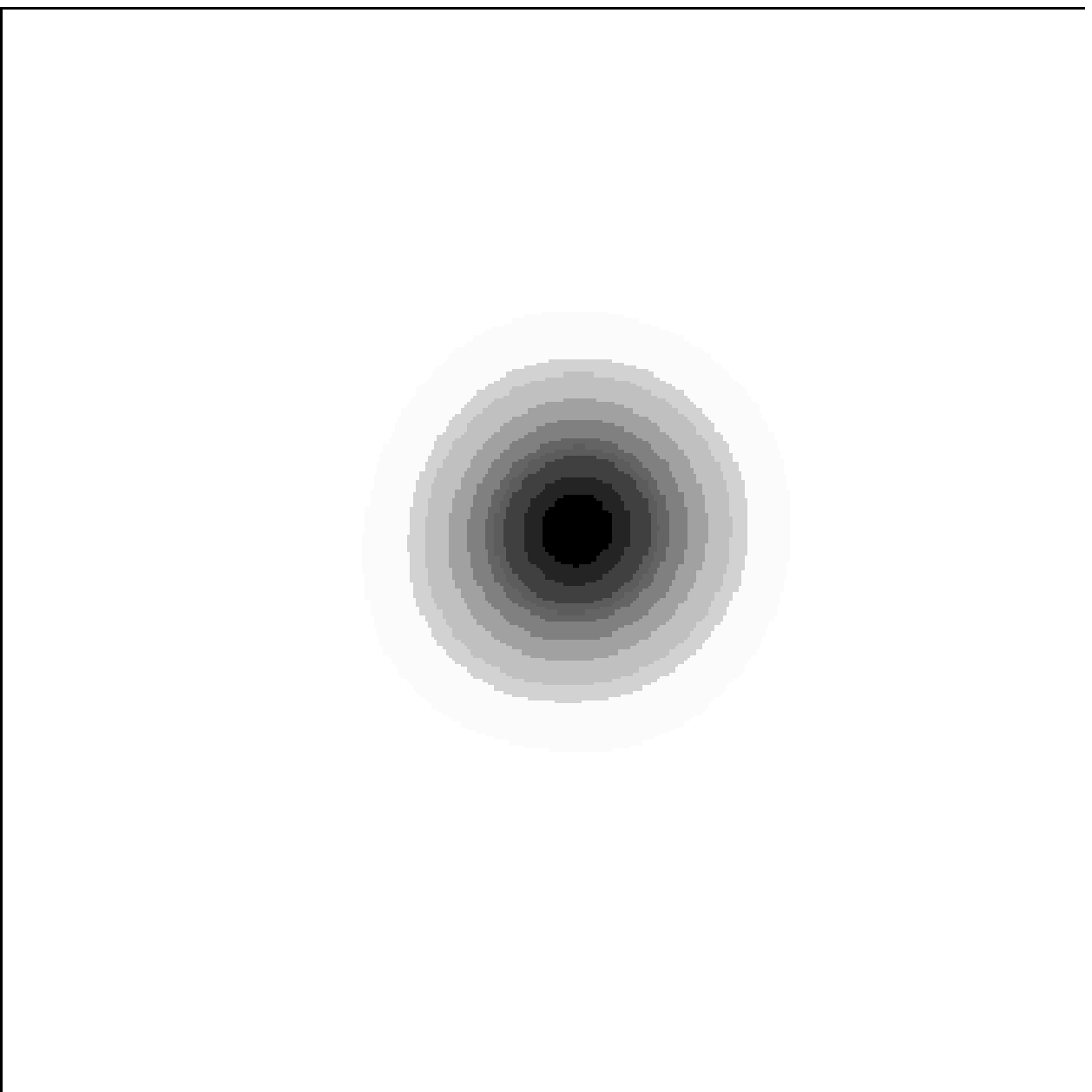}

\caption{Snapshots of the power density distribution in a slice
at the undulator length of $0.5L_\mathrm{g}$ (left plot), $5L_\mathrm{g}$
(middle plot), and $10L_\mathrm{g}$ (right plot).
Calculations have been performed with the simulation code FAST
\cite{fast}.
}

\label{fig:slice-along-z-sase1}
\end{figure}

\begin{figure}[tb]
\includegraphics[width=0.32\textwidth]{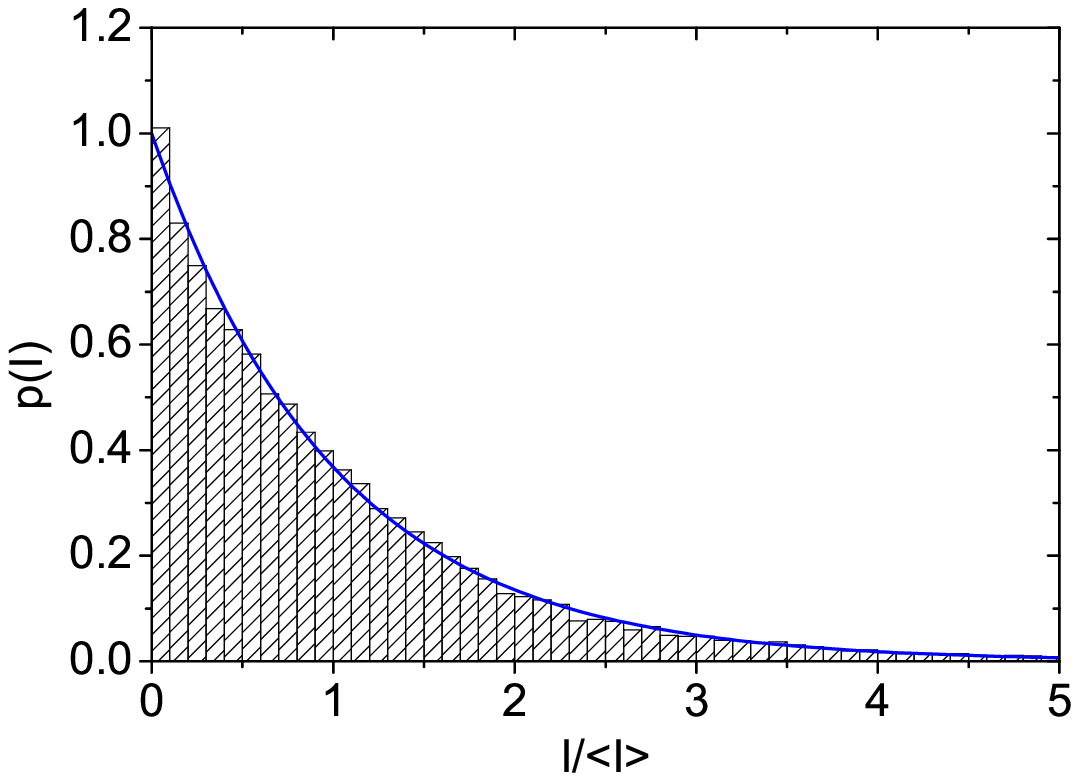}
\includegraphics[width=0.32\textwidth]{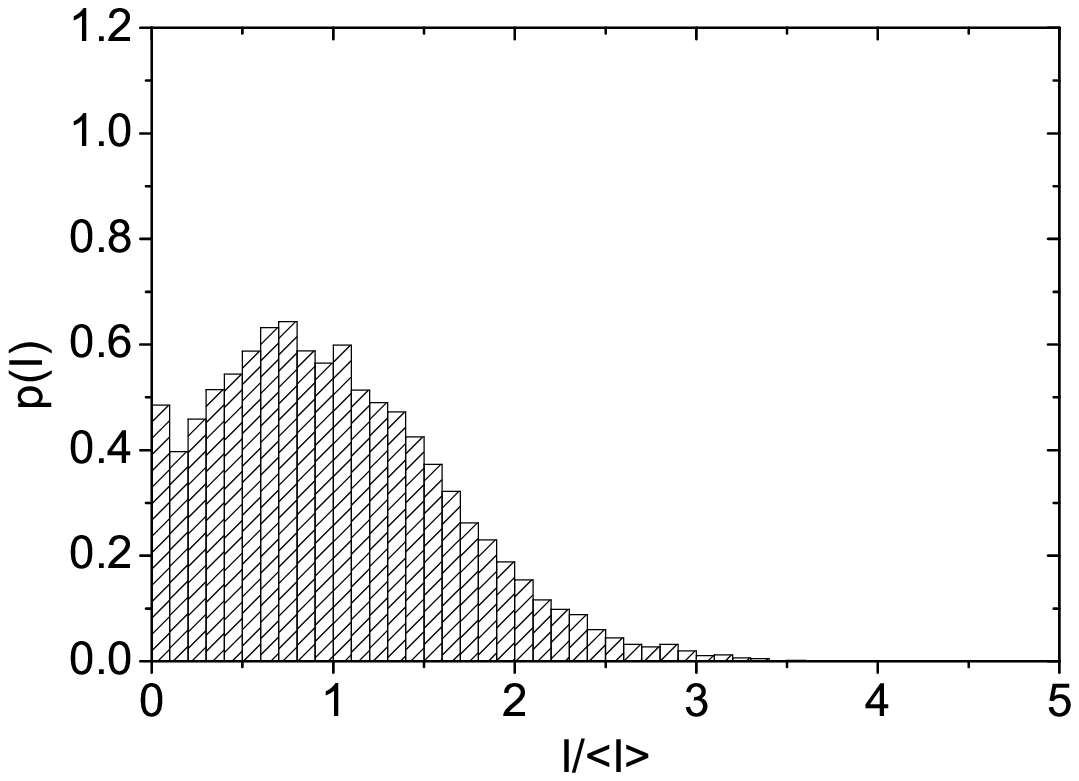}
\includegraphics[width=0.32\textwidth]{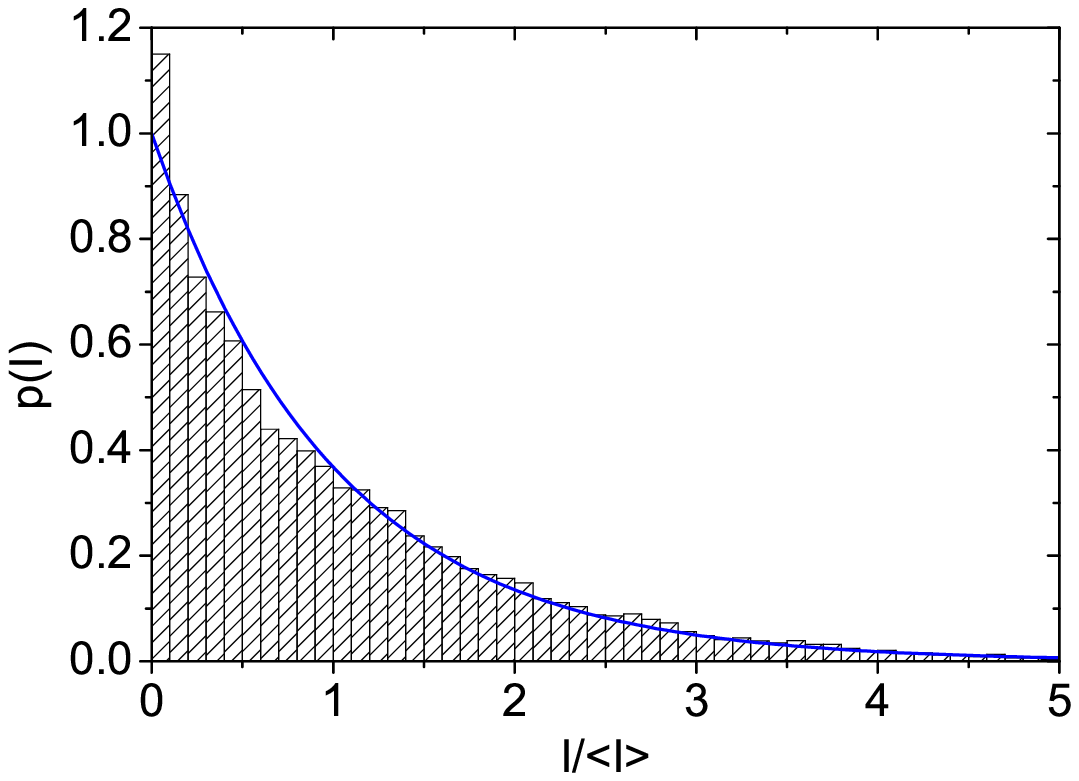}

\includegraphics[width=0.32\textwidth]{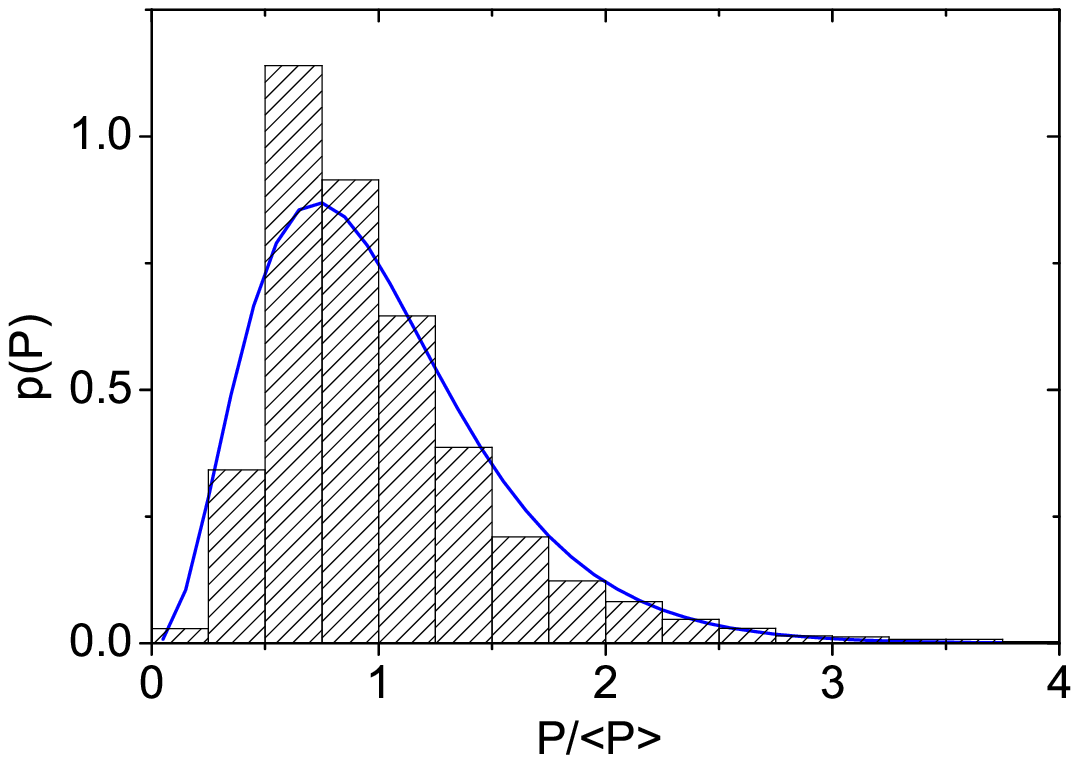}
\includegraphics[width=0.32\textwidth]{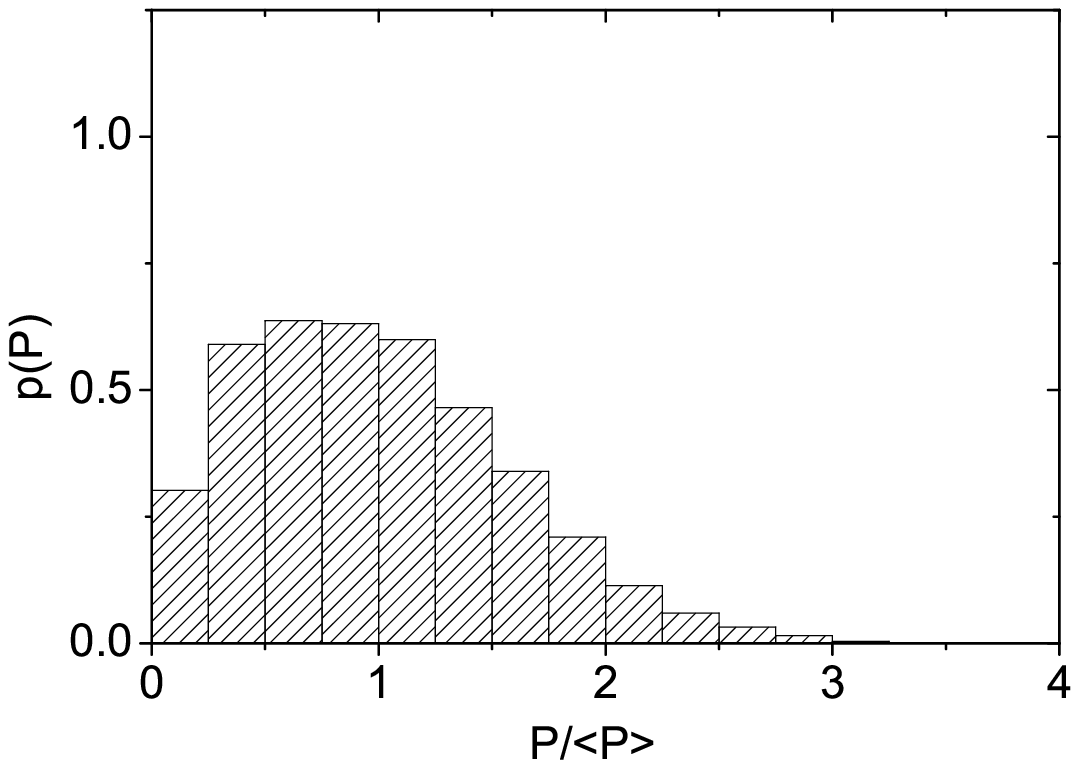}
\includegraphics[width=0.32\textwidth]{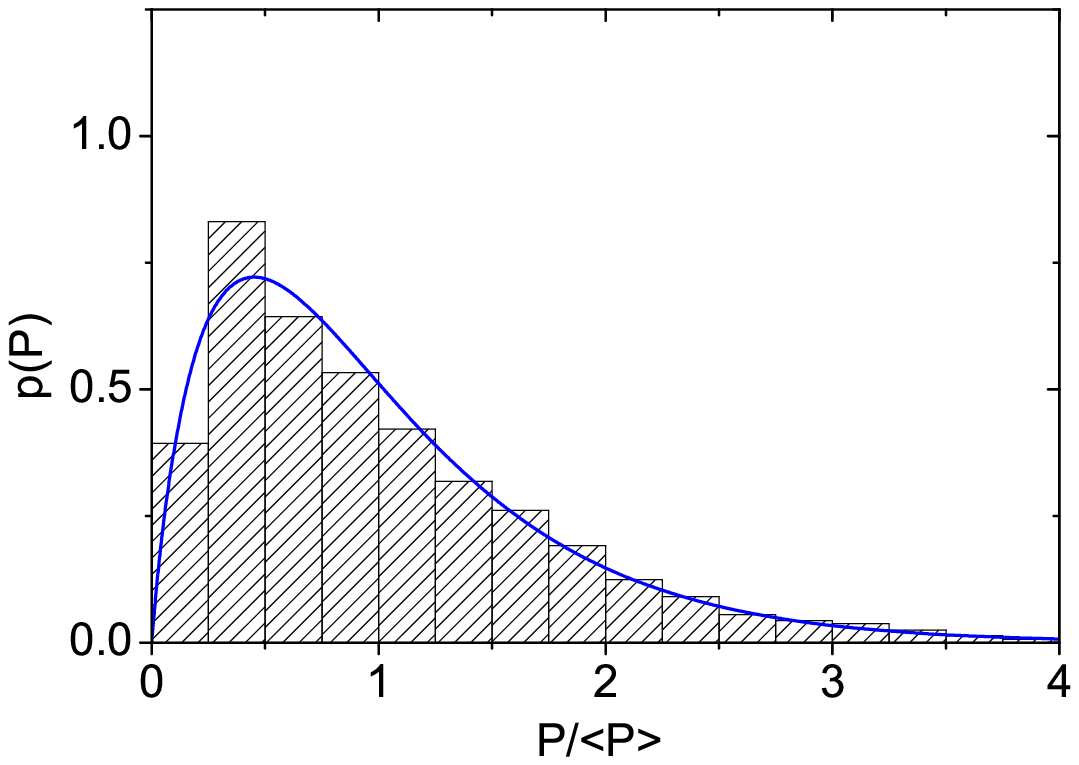}

\caption{
Probability density distributions of the instantaneous power density
$I = |\tilde{E}|^2$ (top), and of the instantaneous radiation power $P$
(bottom)
from
a SASE FEL at different stages of amplification:
linear regime,
saturation regime,
and deep nonlinear regime (undulator length of $5L_\mathrm{g}$,
$10L_\mathrm{g}$, and $15L_\mathrm{g}$, respectively).
Solid lines on the power density histograms (top)
 represent negative exponential distribution
(\ref{neg-exp-1}).
Solid lines on power histograms (bottom)
 represent gamma distribution
(\ref{gamma})
 with $M = 1/\sigma_{\mathrm{P}}^2$.
Here $\hat{\epsilon } = 2$.
Calculations have been performed with the simulation code FAST
\cite{fast}.
}
\label{fig:hprobem2}
\end{figure}

The amplification process in the SASE FEL starts from the shot noise in the
electron beam, then it passes the stage of exponential amplification (high
gain linear stage), and finally enters saturation stage (see
Fig.~\ref{fig:em1-delta}). The field gain length of the fundamental radiation
mode in the high gain linear regime is given by eq.~(\ref{eq:lg}), and
saturation is achieved at the undulator length of about $L_{\mathrm{sat}}
\simeq 10 \times L_{\mathrm{g}}$ for the parameter space of modern X-ray FELs.
Figures \ref{fig:temporal-sase1} and \ref{fig:spectral-sase1} show evolution of
temporal and spectral structure of the radiation pulse along the undulator: at
$0.5 L_{\mathrm{g}}$ (beginning of the undulator), $5 L_{\mathrm{g}}$ (high
gain linear regime), and $10 L_{\mathrm{g}}$ (saturation regime). Figure
\ref{fig:slice-along-z-sase1} shows snapshots of the intensity distributions
across a slice of the photon pulse. We see that many transverse radiation modes
are excited when the electron beam enters the undulator. Radiation field
generated by SASE FEL consists of wavepackets (spikes \cite{1d4}) which
originate from fluctuations of the electron beam density. The typical length of
a spike is about coherence length. Spectrum of the SASE FEL radiation also
exhibits a spiky structure. The spectrum width is inversely proportional to the
coherence time, and a typical width of a spike in a spectrum is inversely
proportional to the pulse duration. The amplification process selects a narrow
band of the radiation, the coherence time increases, and spectrum
shrinks. Transverse coherence is also improved due to the mode selection
process (\ref{eq:modes}).

Figure~\ref{fig:hprobem2} shows the probability distributions of the
instantaneous power density $I \propto |\tilde{E}|^2$ (plots on top), and of
the instantaneous radiation power $P \propto \int I(\vec{r}_{\perp})
\D\vec{r}_{\perp}$ (plots on bottom). We see that transverse and longitudinal
distributions of the radiation intensity exhibit rather chaotic behavior. On
the other hand, probability distributions of the instantaneous power density
$I$ and of the instantaneous radiation power $P$ look more elegant and seem to
be described by simple functions. The origin of this
fundamental simplicity relates to the properties of the electron beam. The shot
noise in the electron beam has a statistical nature that significantly
influences characteristics of the output radiation from a SASE FEL.
Fluctuations of the electron beam current density serve as input signals in a
SASE FEL. These fluctuations always exist in the electron beam due to the
effect of shot noise. Initially fluctuations are not correlated in space and
time, but when the electron beam enters the undulator, beam
modulation at frequencies close to the resonance frequency of the FEL amplifier
initiates the process of the amplification of coherent radiation.

Let us consider microscopic picture of the electron beam current at the
entrance of the undulator. The electron beam current consists of moving
electrons randomly arriving at the entrance of the undulator:

\begin{displaymath}
I(t) = (-e) \sum _{k=1}^N \delta (t - t_k) \ ,
\end{displaymath}

\noindent where $\delta(\ldots )$ is delta-function, (-e) is the charge
of the electron, $N$ is the number of electrons in a bunch and $t_k$ is the
random arrival time of the electron to the undulator entrance. The electron
beam current $I(t)$ and its Fourier transform $\bar{I}(\omega )$ are connected
by Fourier transformations :

\begin{eqnarray}
I(t) & = &
(-e) \sum _{k=1}^N \delta (t - t_k) =
\frac{1}{2\pi } \int \limits_{-\infty }^{\infty } \bar{I}(\omega )
e^{-i\omega t} d\omega
\ ,
\nonumber \\
\bar{I}(\omega ) & = & \int \limits_{-\infty }^{\infty }
e^{i\omega t} I(t) dt = (-e) \sum _{k=1}^N e^{i\omega t_k} \ .
\label{eq:shot-fourier}
\end{eqnarray}

\noindent It follows from eq.~(\ref{eq:shot-fourier}) that the Fourier
transformation of the input current, $\bar{I}(\omega )$, is the sum of large
number of complex phasors with random phases $\phi_k = \omega t_k$. Thus,
harmonics of the electron beam current are described with gaussian statistics.

The FEL process is just an amplification of the initial shot noise in the
narrow band near the resonance wavelength $\lambda $ when both harmonics of the
beam current and radiation are growing. An FEL amplifier operating in the
linear regime is just a linear filter, and the Fourier harmonic of the
radiation field is simply proportional to the Fourier harmonic of the electron
beam current, $\bar{E}(\omega) = H_A(\omega-\omega_0) \bar{I}(\omega)$. Thus,
the statistics of the radiation are gaussian -- the same as of the shot noise
in the electron beam. This kind of radiation is usually referred to as
completely chaotic polarized light, a well known object in the field of
statistical optics \cite{goodman}. For instance, the higher order correlation
functions (time and spectral) are expressed via the first order correlation
function. The spectral density of the radiation energy and the first-order time
correlation function form a Fourier transform pair (Wiener Khintchine theorem).
The real and imaginary parts of the slowly varying complex amplitudes of the
electric field of the electromagnetic wave, $\tilde{E}$ , have a Gaussian
distribution. The instantaneous power density, $I = |\tilde{E}|^2$, fluctuates
in accordance with the negative exponential distribution (see
Fig.~\ref{fig:hprobem2}):

\begin{equation}
p(I) = \frac{1}
{\langle I \rangle }
\exp\left(-\frac{I}
{\langle I \rangle }\right) \ .
\label{neg-exp-1}
\end{equation}

\noindent Any integral of the power density, like radiation power $P$,
fluctuates in accordance with the gamma distribution:

\begin{equation}
p({P}) = \frac{M^M}{\Gamma (M)}
\left( \frac{P}{\langle P \rangle }\right)^{M-1}
\frac{1}{\langle P \rangle } \exp \left( -M \frac{P}
{\langle P \rangle } \right) \ ,
\label{gamma}
\end{equation}

\noindent where $\Gamma (M)$ is the gamma function with argument $M =
1/\sigma_{\mathrm{P}}^2$, and $\sigma_{\mathrm{P}}^2 = \langle (P -\langle P
\rangle )^2 \rangle / \langle P \rangle^2 $ is the relative dispersion of the
radiation power. For completely chaotic polarized light parameter $M$ has a clear
physical interpretation -- it is the number of modes \cite{book}. Thus, the
relative dispersion of the radiation power directly relates to the coherence
properties of the SASE FEL operating in the linear regime. The degree of
transverse coherence in this case can be defined as \cite{book}:

\begin{equation}
\zeta  = \frac{1}{M} = \sigma^2_{\mathrm{P}} \ .
\label{eq:degcoh-m}
\end{equation}

\noindent It is shown in ref.~\cite{coherence-oc} that such a definition for
the degree of transverse coherence is mathematically equivalent to
(\ref{eq:def-degcoh}).

When amplification process enters nonlinear stage and reaches saturation,
statistics of the radiation significantly deviate from gaussian. Particular
signature of this change is illustrated in Fig.~\ref{fig:hprobem2}. We see that
the probability distribution of the radiation intensity is not the negative
exponential, and the probability distribution of the radiation power visibly
deviates from gamma distribution. Up to now there is no analytical description
of the statistics in the saturation regime, and we refer the reader to the
analysis of the results of numerical simulations \cite{coherence-oc}. General
feature of the saturation regime is that fluctuations of the radiation
intensity are significantly suppressed. We also find that the definition of the
degree of transverse coherence (\ref{eq:degcoh-m}) has no physical sense near
the saturation point.

When we trace the amplification process further in the nonlinear regime, we
obtain that fluctuations of the radiation intensity and radiation power
increase, and relevant probability distributions tend to those given by eqs.
(\ref{neg-exp-1}) and (\ref{gamma}). This behavior hints that the properties of
the radiation from a SASE FEL operating in the deep nonlinear regime tend to be
those of completely chaotic polarized light \cite{1d7,coherence-oc}.
For the deep nonlinear regime we find that the degree of transverse coherence
defined by (\ref{eq:def-degcoh}) again tends to be an agreement with
(\ref{eq:degcoh-m}).

Another practical problem refers to the probability distributions of the
radiation intensity in the frequency domain, like that filtered by
monochromator. For SASE FEL radiation produced in the linear regime the
probability distribution radiation intensity is defined by gaussian statistics,
and it is the negative exponential for a narrow band monochromator. When
amplification process enters saturation regime, this property still holds for
the case of long electron pulse \cite{book,1d7}, and is violated significantly
for the case of short electron bunch, of about or less than coherence length.
In the latter case fluctuations of the radiation intensity after narrow band
monochromator are significantly suppressed as it has been predicted
theoretically and measured experimentally at the free electron laser FLASH at
DESY operating in a femtosecond mode \cite{ssy-sb,flash-statistics}.

All considerations presented above are related to the fundamental harmonic of
the SASE FEL radiation. Radiation from a SASE FEL with a planar undulator has
rich harmonics contents. Intensities of even harmonics are suppressed
\cite{sase-2nd-harm}, but odd harmonics provide significant contribution to the
total radiation power \cite{hg-1,hg-2,hg-2a,hg-3,kim-1}. Comprehensive studies
of the statistical properties of the odd harmonics have been performed in paper
\cite{3rd-harm-prstab}. It has been found  that the statistics of the
high-harmonic radiation from the SASE FEL changes significantly with respect to
the fundamental harmonic (with respect to gaussian statistics). For the
fundamental harmonic the probability density function of the intensity is the
negative exponential distribution: $p(W) =
\langle{W}\rangle^{-1}\exp(-W/\langle{W}\rangle)$. Mechanism of the higher
harmonic generation is equivalent to the transformation of the intensity $W$ as
$z = (W)^{n}$, where $n$ is the harmonic number. It has been shown in
\cite{3rd-harm-prstab} that the probability distribution for the intensity of
the n-th harmonic is given by:

\begin{equation}
p(z) =
\frac{z}{n\langle{W}\rangle}
{z}^{(1-n)/n}\exp(-z^{1/n}/\langle{W}\rangle)
\ .
\label{eq:prob-hg}
\end{equation}

\noindent The expression for the mean value is $\langle{z}\rangle =
n!\langle{W}\rangle^{n}$. Thus, the $n$th-harmonic radiation for the SASE FEL
has an intensity level roughly $n!$ times larger than the corresponding
steady-state case, but with more shot-to-shot fluctuations compared to the
fundamental \cite{kim-1}. Nontrivial behavior of the intensity of the high
harmonic reflects the complicated nonlinear transformation of the fundamental
harmonic statistics. In this case a gaussian statistics are no longer valid.
Practically this behavior occurs only in the very end of high gain exponential
regime when coherent radiation intensity exceeds an incoherent one. When
amplification enters nonlinear stage, probability distributions change
dramatically on a scale of the gain length, and already in the saturation
regime (and further downstream the undulator) the probability distributions of
the radiation intensity of higher harmonics are pretty much close to the
negative exponential distribution \cite{3rd-harm-prstab}.

\section{Characteristics of the radiation from SASE FEL operating in the
saturation regime}

\begin{figure}[tb]
\begin{center}
\epsfig{file=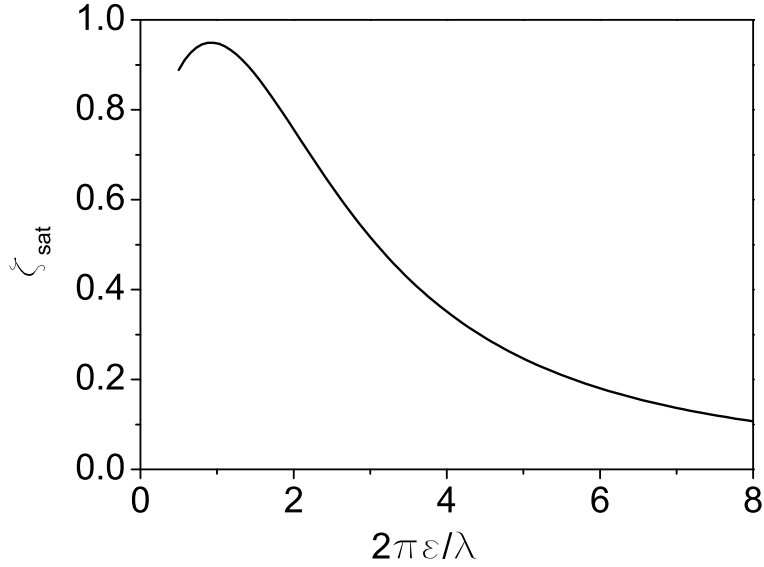,width=0.75\textwidth}
\end{center}
\caption{
Degree of transverse coherence $\zeta _{\mathrm{sat}}$ in the saturation point
versus parameter $\hat{\epsilon } $. The number of electrons in the coherence
volume is $N_\mathrm{c} = 4 \times 10^{6}$.
Calculations have been performed with the simulation code FAST
\cite{fast}.
}
\label{fig:bril-dcoh}
\end{figure}

In Fig.~\ref{fig:em1-delta} we present evolution of the main characteristics of
a SASE FEL along the undulator. If one traces evolution of the brilliance
(degeneracy parameter) of the radiation along the undulator length,  there is
always the point (defined as the saturation point \cite{coherence-oc})
where the brilliance reaches maximum value. The best properties of the
radiation in terms of transverse and longitudinal coherence are reached just
before the saturation point, and then degrade significantly despite the
radiation power continuing to grow with the undulator length.

Application of similarity techniques allows us to derive universal parametric
dependencies of the output characteristics of the radiation at the saturation
point. As we mentioned in Section 2, within accepted approximations (optimized
SASE FEL and negligibly small energy spread in the electron beam), normalized
output characteristics of a SASE FEL at the saturation point are functions of
only two parameters: $\hat{\epsilon} = 2\pi \epsilon / \lambda $ and the number
of electrons in the volume of coherence $N_\mathrm{c} = I N_\mathrm{g} \lambda
/c$, where $N_\mathrm{g} = L_\mathrm{g} / \lambda _\mathrm{w}$ is the number of
undulator periods per gain length. Characteristics of practical interest are:
saturation length $L_\mathrm{sat}$, saturation efficiency $\eta _\mathrm{sat} =
P_\mathrm{sat} / P_\mathrm{b}$ (ratio of the radiation power to the electron
beam power $P_\mathrm{b} = \gamma mc^2 I/e$), coherence time $\tau_\mathrm{c}$,
degree of transverse coherence $\zeta $, degeneracy parameter $\delta $, and
brilliance $B_\mathrm{r}$.
Applications of similarity techniques to the results of
numerical simulations of a SASE FEL \cite{coherence-oc} gives us the following
result:

\begin{eqnarray}
\hat{L}_\mathrm{sat} & = & \Gamma L_\mathrm{sat} \simeq
2.5 \times \hat{\epsilon }^{5/6} \times \ln N_\mathrm{c} \ ,
\nonumber \\
\hat{\eta } & = & P / (\bar{\rho } P_\mathrm{b}) \simeq
0.17 /\hat{\epsilon } \ ,
\nonumber \\
\hat{\tau_\mathrm{c}} & = & \bar{\rho } \omega \tau_\mathrm{c}
\simeq 1.16 \times \sqrt{\ln N_\mathrm{c}} \times \hat{\epsilon }^{5/6} \ ,
\nonumber \\
\sigma _{\omega } & = & \sqrt{\pi }/\tau_\mathrm{c} \ .
\label{eq:parameters-sat-1}
\end{eqnarray}

\noindent
These expressions provide reasonable practical
accuracy for $\hat{\epsilon } \gtrsim 0.5$.
With logarithmic accuracy
in terms of $N_{\mathrm{c}}$ characteristics of the SASE FEL expressed in a
normalized form are functions of the only parameter $\hat{\epsilon}$.
The saturation length, FEL efficiency, and coherence time exhibit monotonous
behavior
in the parameter space of modern XFELs ($\hat{\epsilon } \simeq 0.5 \ldots 5$).
Situation is a bit complicated with the degree of transverse
coherence as one can see in Fig.~\ref{fig:bril-dcoh}. The degree of transverse
coherence reaches a maximum value in the range of $\hat{\epsilon } \sim 1$, and
drops at small and large values of $\hat{\epsilon }$. At small values of the
emittance, the degree of transverse coherence is limited by the interdependence
of poor longitudinal coherence and transverse coherence \cite{trcoh-oc}. Due to
the start-up from shot noise, every radiation mode entering
eq.~(\ref{eq:modes}) is excited within finite spectral bandwidth. This means
that the radiation from a SASE FEL is formed by
many fundamental TEM$_{00}$ modes with different frequencies. The transverse
distribution of the radiation field of the mode is also different for different
frequencies. Smaller values of $\hat{\epsilon }$ (smaller value of the
diffraction parameter) correspond to larger frequency bandwidths. This
effect explains the decrease of the transverse coherence at small values of
$\hat{\epsilon }$. The degree of transverse coherence asymptotically approaches
unity as $(1 -\zeta) \propto 1/z \propto 1/\ln N_\mathrm{c}$ at small values of
the emittance.

In the case of large emittance the degree of transverse coherence is defined by
the contents of higher transverse modes \cite{coherence-oc,coherence-anal-oc}.
When $\hat{\epsilon }$ increases, the diffraction parameter
increases as well, leading to the degeneration of the radiation modes
\cite{book}. The amplification process in the SASE FEL passes limited number of
the field gain lengths, and starting from some value of $\hat{\epsilon }$, the
linear stage of amplification becomes too short to provide a mode selection
process (\ref{eq:modes}). When the amplification process enters the nonlinear
stage, the mode content of the radiation becomes richer due to
independent growth of the radiation modes in the nonlinear medium. Thus, at
large values of $\hat{\epsilon }$ the degree of transverse coherence is limited
by poor mode selection. The degree of
transverse coherence scales as $\zeta _\mathrm{sat} \propto (\ln
N_\mathrm{c}/\hat{\epsilon })^{2}$ in the asymptote of large emittance. To
avoid complications, we present here just a fit for the degree of transverse
coherence for the number of electrons in the coherence volume $N_\mathrm{c} = 4
\times 10^{6}$:

\begin{equation}
\zeta _\mathrm{sat} \simeq \frac{1.1 \hat{\epsilon }^{1/4}}{1+0.15
\hat{\epsilon }^{9/4}} \ .
\label{eq:delta-sat}
\label{eq:parameters-sat-2}
\end{equation}

\noindent Recalculation from reduced to dimensional parameters is
straightforward. For instance, saturation length is $L_\mathrm{sat} \simeq 0.6
\times L_\mathrm{g} \times \ln N_\mathrm{c}$. Using (\ref{eq:parameters-sat-1})
and (\ref{eq:parameters-sat-2}) we can calculate normalized degeneracy
parameter $\hat{\delta } = \hat{\eta }\zeta \hat{\tau_\mathrm{c}}$ and then the
brilliance (\ref{eq:bril-1}):

\begin{equation}
B_r \left[ \frac{\mathrm{photons}}
{\mathrm{sec \ mrad^2 \ mm^2 \ 0.1\% \ bandw.)}}
\right] \simeq
4.5\times 10^{31} \times
\frac{I [\mathrm{kA}] \times
E [\mathrm{GeV}]}{\lambda [\mathrm{\AA } \ ]}
\times \hat{\delta } \ .
\label{eq:bril-pract}
\end{equation}

Properties of the odd harmonics of the radiation from a SASE FEL with a
planar undulator operating in the saturation regime also possess simple
features. In the case of cold electron beam contributions of the higher odd
harmonics to the FEL power are functions of the only undulator parameter $K$
\cite{3rd-harm-prstab}:

\begin{equation}
\frac{\langle W_3 \rangle}{\langle W_1 \rangle} \vert _{\mathrm{sat}} =
0.094 \times \frac{K_3^2}{K_1^2} \ , \qquad
\frac{\langle W_5 \rangle}{\langle W_1 \rangle} \vert _{\mathrm{sat}}=
0.03 \times \frac{K_5^2}{K_1^2} \ .
\label{eq:sat35}
\end{equation}

\noindent Here $K_h = K(-1)^{(h-1)/2} [J_{(h-1)/2}(Q) - J_{(h+1)/2}(Q)]$, $Q =
K^2/[2(1+K^2)]$, and $h$ is an odd integer. Influence of the energy spread and
emittance leads to significant decrease of the power of higher harmonics, up to
a factor of three for the third harmonic, and a factor of up to ten for the fifth
harmonic. Power of the higher harmonics is subjected
to larger fluctuations than the power of the fundamental harmonic as we
mentioned in the previous section. The coherence time at saturation scales
inversely proportional to the harmonic number, while relative spectrum
bandwidth remains constant with the harmonic number.

\section{Estimations in the framework of
the one-dimensional model}

An estimation of SASE FEL characteristics is frequently performed in the
framework of the one-dimensional model in terms of the FEL parameter $\rho
$ \cite{bon-rho}:

\begin{equation}
\rho  = \frac{\lambda_{\mathrm{w}}}{4 \pi }
\left[ \frac
{4\pi ^2 j_0 K^2 A_{\mathrm{JJ}}^2}
{I_{A} \lambda_{\mathrm{w}} \gamma^3}
\right]^{1/3} \ ,
\label{eq:rho-1D}
\end{equation}

\noindent where $j_0 = I/(2\pi \sigma ^2)$ is the beam current density,
$\sigma = \sqrt{\beta \epsilon _{\mathrm{n}}/\gamma}$ is rms
transverse size of the electron beam, and $\beta $ is external focusing beta
function. FEL parameter $\rho $ relates to the efficiency
parameter of the 3D FEL theory as $\rho = \bar{\rho}/B^{1/3}$.
Basic characteristics of the SASE FEL are estimated in terms of the
parameter $\rho $ and number of cooperating electrons
$N_{\mathrm{c}} = I/(e\rho \omega )$. Here we present a set of
simple formulae extracted from \cite{book,1d4,1d7}:

\begin{eqnarray}
\mathrm{The \ field \ gain \ length:} \quad
&\mbox{}&
L_{\mathrm{g}} \sim
\frac{\lambda _{\mathrm{w}}}{4\pi \rho } \ ,
\nonumber \\
\mathrm{Saturation \ length:} \quad
&\mbox{}&
L_{\mathrm{sat}} \sim
\frac{\lambda _{\mathrm{w}}}{4\pi \rho }
\left[
3 + \frac{\ln N_\mathrm{c}}{\sqrt{3}}
\right]
\nonumber \\
\mathrm{Effective \ power \ of \ shot \ noise:} \quad
&\mbox{}&
\frac{P_{\mathrm{sh}}}{\rho P_{\mathrm{b}}} \simeq
\frac{3}{ N_{\mathrm{c}} \sqrt{ \pi \ln N_{\mathrm{c}} }} \ ,
\nonumber \\
\mathrm{Saturation \ efficiency:} \quad
&\mbox{}&
\rho  \ ,
\nonumber \\
\mathrm{The \ power \ gain \ at \ saturation:} \quad
&\mbox{}&
G \simeq
\frac{1}{3} N_{\mathrm{c}} \sqrt{ \pi \ln N_{\mathrm{c}} } \ ,
\nonumber \\
\mathrm{Coherence \ time \ at \ saturation:} \quad
&\mbox{}&
\tau _{\mathrm{c}} \simeq
\frac{1}{\rho \omega }
\sqrt{ \frac{\pi \ln N_{\mathrm{c}} }{18} } \ .
\nonumber \\
\mathrm{Spectrum \ bandwidth:} \quad
&\mbox{}&
\sigma _{\omega }  =  \sqrt{\pi }/\tau_\mathrm{c} \ ,
\label{eq:1D-SASE}
\end{eqnarray}

\noindent In many cases this set of formulas can help quickly estimate main
parameters of a SASE FEL but it does not provide complete self-consistent basis
for optimization of this device.

\section*{Acknowledgments}

We are grateful to Dr. Pavle Jurani\'c for careful reading of the
manuscript.

\end{document}